\documentclass[journal]{IEEEtran}%
\usepackage[dvips]{epsfig}
\usepackage{graphicx}
\usepackage{amsmath}%
\usepackage{amsfonts}%
\usepackage{amssymb}
\ifCLASSINFOpdf
\else
\fi

\begin{document}

\title{Accuracy of Transfer Matrix Approaches for Solving the Effective Mass
Schr\"{o}dinger Equation}
\author{Christian Jirauschek~\IEEEmembership{Member,~IEEE}
\linebreak(Dated: 16 June 2011, published as IEEE J. Quantum Electron. 45, 1059–-1067 (2009))
\linebreak \linebreak \copyright 2009 IEEE. Personal use of this material is permitted. Permission from IEEE must be obtained for all other uses, in any current or future media, including reprinting/republishing this material for advertising or promotional purposes, creating new collective works, for resale or redistribution to servers or lists, or reuse of any copyrighted component of this work in other works.
 \thanks{This work is supported by the Emmy Noether program of the German Research
Foundation (DFG, JI115/1-1).} \thanks{C. Jirauschek is with the Institute for
Nanoelectronics, TU M\"{u}nchen, Arcisstr. 21, D-80333 M\"{u}nchen, Germany;
e-mail: jirauschek@tum.de.}}

\maketitle

\begin{abstract}
The accuracy of different transfer matrix approaches, widely used to solve the
stationary effective mass Schr\"{o}dinger equation for arbitrary
one-dimensional potentials, is investigated analytically and numerically. Both
the case of a constant and a position dependent effective mass are considered.
Comparisons with a finite difference method are also performed. Based on
analytical model potentials as well as self-consistent Schr\"{o}dinger-Poisson
simulations of a heterostructure device, it is shown that a symmetrized
transfer matrix approach yields a similar accuracy as the Airy function method
at a significantly reduced numerical cost, moreover avoiding the numerical
problems associated with Airy functions.

\end{abstract}

\markboth{Accuracy of Transfer Matrix Approaches for Solving the Effective
Mass Schr\"{o}dinger Equation}{Jirauschek}%

\begin{IEEEkeywords}
Quantum effect semiconductor devices, Quantum well devices, Quantum theory,
Semiconductor heterojunctions, Eigenvalues and eigenfunctions, Numerical
analysis, Tunneling, MOS devices
\end{IEEEkeywords}                 

\IEEEpeerreviewmaketitle

\section{Introduction}

\label{Intro}

\IEEEPARstart{T}{ransfer} matrix methods provide an important tool for
investigating bound and scattering states in quantum structures. They are
mainly used to solve the one-dimensional Schr\"{o}dinger or effective mass
equation, e.g., to obtain the quantized eigenenergies in quantum well
heterostructures and metal-oxide-semiconductor structures or the transmission
coefficient of potential barriers
\cite{1987JAP....61.1497A,1990IJQE...26.2025J,2000JAP....87.7931C,1997plds.book.....D}%
. Analytical expressions for the transfer matrices are only available in
certain cases, as for constant or linear potential sections and potential
steps \cite{1997plds.book.....D}. An arbitrary potential can then be treated
by approximating it for example in terms of piecewise constant or linear
segments, for which analytical transfer matrices exist. For constant potential
segments, the matrices are based on complex exponentials
\cite{1987JAP....61.1497A,1990IJQE...26.2025J}, while the linear potential
approximation requires the evaluation of Airy functions
\cite{1990IJQE...26.2025J}.

Many applications call for highly accurate methods, e.g., quantum cascade
laser structures where layer thickness changes by a few \AA\ already lead to
significantly modified wavefunctions, resulting in altered device properties
\cite{2005ApPhL..86k1115V,2007JAP...101h6109J}. Also numerical efficiency is
crucial, especially in cases where the Schr\"{o}dinger equation has to be
solved repeatedly. Examples are the shooting method where the eigenenergies of
bound states are found by energy scans, or Schr\"{o}dinger-Poisson solvers
working in an iterative manner \cite{2000JAP....87.7931C}. Besides providing
accurate results at moderate computational cost, an algorithm is expected to
be numerically robust, and a straightforward implementation is also advantageous.

Besides transfer matrices, also other methods are frequently used, in
particular finite difference or finite element schemes
\cite{1998hqd.book.....F,1990PhRvB..4112047J}. For scattering state
calculations, they are complimented by suitable transparent boundary
conditions, resulting in the Quantum Transmitting Boundary Method (QTBM)
\cite{1998hqd.book.....F,1990JAP....67.6353L}. The transfer matrix method
tends to be less numerically stable than the QTBM, since for multiple or
extended barriers, numerical instabilities can arise due to an exponential
blowup caused by roundoff errors \cite{1998hqd.book.....F}. This issue can
however be overcome, for example by using a somewhat modified matrix approach,
the scattering matrix method \cite{1988PhRvB..38.9945K}. In this case, the
transfer matrices of the individual segments are not used to compute the
overall transfer matrix, but rather the scattering matrix of the structure. In
addition, transfer matrices have many practical properties, such as their
intuitiveness particularly for scattering states, the intrinsic current
conservation, and the exact treatment of potential steps, which arise at the
interfaces of differing materials. This makes them especially suitable and
popular for 1-D heterostructures or metal-oxide-semiconductor
structures, providing a simple, accurate and efficient simulation method
\cite{1990IJQE...26.2025J}.

As mentioned above, transfer matrices are usually based on a piecewise
constant or piecewise linear approximation of an arbitrary potential, giving
rise to exponential and Airy function solutions, respectively. The main
strength of the Airy function approach is that it provides an exact solution
for structures consisting of piecewise linear potentials, and hence only
requires few segments for approximating almost linear potentials with
sufficient accuracy. On the other hand, Airy functions are much more
computationally demanding than exponentials, and also prone to numerical
overflow for regions with nearly flat potential \cite{1996IJQE...32.1093V}.
Thus, great care has to be taken to avoid these problems, and to evaluate the
Airy functions in an efficient way \cite{2001JAP....90.6120D}.

It would be desirable to combine the advantages of both methods, namely the
accuracy of the piecewise linear approximation and the computational
convenience of the exponential transfer matrix scheme. In this paper, we
evaluate the accuracy and efficiency of the different transfer matrix
approaches, taking into account both bound and scattering states. In this
context, analytical expressions for the corresponding local discretization
error are derived. We furthermore evaluate the different approaches
numerically on the basis of an analytically solvable model potential, and also
draw comparisons to the QTBM. In particular, we demonstrate that a symmetrized
exponential matrix approach is able to provide an accuracy comparable to that
of the Airy function method, without having its problems and drawbacks. In our
investigation, we will consider both the case of a constant effective mass and
the more general case of a position dependent effective mass.

\section{Transfer matrix approach}

\label{sec:tra}

\begin{figure}[t]
\centering\includegraphics[width=3.5in]{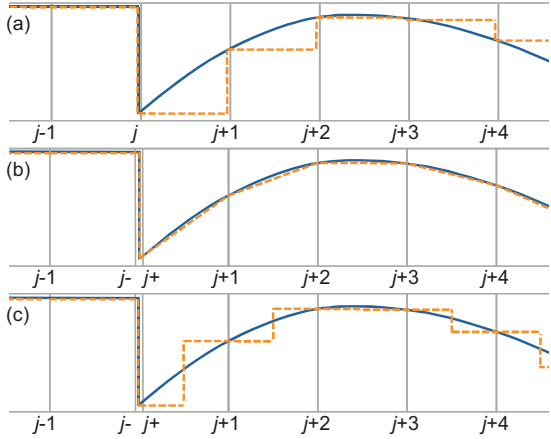}\caption{Various transfer
matrix schemes applied to segmented potential. Shown is the exact (solid line)
and approximated (dashed line) potential. (a) Piecewise constant potential
approximation. (b) Piecewise linear approximation. (c) Piecewise contant
approximation for symmetrized transfer matrix.}%
\label{transfer}%
\end{figure}

In a single-band approximation, the wavefunction $\psi$ of an electron with
energy $E$ in a one-dimensional quantum structure can be described by the
effective mass equation
\begin{equation}
\left[  -\frac{\hbar^{2}}{2}\partial_{z}\frac{1}{m^{\ast}\left(  z\right)
}\partial_{z}+V(z)-E\right]  \psi(z)=0. \label{effm}%
\end{equation}
Here, the effective mass $m^{\ast}$ and the potential $V$ generally depend on
the position $z$ in the structure. For applying the transfer matrix scheme, we
divide the structure into segments, see Fig. \ref{transfer}, which can vary in
length. Potential and effective mass discontinuities can be treated exactly in
transfer matrix approaches by applying corresponding matching conditions. To
take advantage of this fact and obtain optimum accuracy, the segments should
be chosen so that band edge discontinuities, as introduced by heterostructure
interfaces, do not lie within a segment, but rather at the border between two segments.

\subsection{Conventional transfer matrices}

\label{sub:tra1}

For the piecewise constant potential approach (Fig. \ref{transfer}(a)), the
potential and effective mass in each segment $j$ are approximated by constant
values, e.g., $V_{j}=V\left(  z_{j}\right)  $, $m_{j}^{\ast}=m^{\ast}\left(
z_{j}\right)  $ for $z_{j}\leq z<z_{j}+\Delta_{j}=z_{j+1}$, and a jump
$V_{j}\rightarrow V_{j+1}$, $m_{j}^{\ast}\rightarrow m_{j+1}^{\ast}$ at the
end of the segment \cite{1990IJQE...26.2025J}. The solution of (\ref{effm}) is
for $z_{j}\leq z<z_{j+1}$ then given by%
\begin{equation}
\psi\left(  z\right)  =A_{j}\exp\left[  \mathrm{i}k_{j}\left(  z-z_{j}\right)
\right]  +B_{j}\exp\left[  -\mathrm{i}k_{j}\left(  z-z_{j}\right)  \right]  ,
\label{psi}%
\end{equation}
where $k_{j}=\sqrt{2m_{j}^{\ast}\left(  E-V_{j}\right)  }/\hbar$ is the
wavenumber (for $E<V_{j}$, we obtain $k_{j}=\mathrm{i}\kappa_{j}%
=\mathrm{i}\sqrt{2m_{j}^{\ast}\left(  V_{j}-E\right)  }/\hbar$)
\cite{1990IJQE...26.2025J}. The matching conditions for the wavefunction at
the potential step read%
\begin{align}
\psi\left(  z_{0}+\right)   &  =\psi\left(  z_{0}-\right)  ,\nonumber\\
\left[  \partial_{z}\psi\left(  z_{0}+\right)  \right]  /m^{\ast}\left(
z_{0}+\right)   &  =\left[  \partial_{z}\psi\left(  z_{0}-\right)  \right]
/m^{\ast}\left(  z_{0}-\right)  , \label{match}%
\end{align}
where $z_{0}+$ and $z_{0}-$ denote the positions directly to the right and
left of the step, here located at $z_{0}=z_{j+1}$ \cite{1997plds.book.....D}.
The amplitudes $A_{j+1}$ and $B_{j+1}$ are related to $A_{j}$ and $B_{j}$ by%
\begin{equation}
\left(
\begin{array}
[c]{c}%
A_{j+1}\\
B_{j+1}%
\end{array}
\right)  =T_{j,j+1}\left(
\begin{array}
[c]{c}%
A_{j}\\
B_{j}%
\end{array}
\right)  , \label{mat}%
\end{equation}
with the transfer matrix
\begin{align}
T_{j,j+1}  &  =T_{j\rightarrow j+1}T_{j}\left(  \Delta_{j}\right) \nonumber\\
&  =\left(
\begin{array}
[c]{cc}%
\frac{\beta_{j+1}+\beta_{j}}{2\beta_{j+1}}e^{\mathrm{i}k_{j}\Delta_{j}} &
\frac{\beta_{j+1}-\beta_{j}}{2\beta_{j+1}}e^{-\mathrm{i}k_{j}\Delta_{j}}\\
\frac{\beta_{j+1}-\beta_{j}}{2\beta_{j+1}}e^{\mathrm{i}k_{j}\Delta_{j}} &
\frac{\beta_{j+1}+\beta_{j}}{2\beta_{j+1}}e^{-\mathrm{i}k_{j}\Delta_{j}}%
\end{array}
\right)  . \label{T1}%
\end{align}
Equation (\ref{T1}) is the product of the transfer matrix for a flat potential%
\begin{equation}
T_{j}\left(  \Delta_{j}\right)  =\left(
\begin{array}
[c]{cc}%
e^{\mathrm{i}k_{j}\Delta_{j}} & 0\\
0 & e^{-\mathrm{i}k_{j}\Delta_{j}}%
\end{array}
\right)  ,
\end{equation}
obtained from (\ref{psi}), and the potential step matrix%
\begin{equation}
T_{j\rightarrow j+1}=\frac{1}{2\beta_{j+1}}\left(
\begin{array}
[c]{cc}%
\beta_{j+1}+\beta_{j} & \beta_{j+1}-\beta_{j}\\
\beta_{j+1}-\beta_{j} & \beta_{j+1}+\beta_{j}%
\end{array}
\right)  \label{Tst}%
\end{equation}
with $\beta_{j}=k_{j}/m_{j}^{\ast}$, derived from (\ref{match})
\cite{1997plds.book.....D}. The relation between the amplitudes at the left
and right boundaries of the structure, $A_{0},B_{0}$ and $A_{N},B_{N}$, can be
obtained from%
\begin{align}
\left(
\begin{array}
[c]{c}%
A_{N}\\
B_{N}%
\end{array}
\right)   &  =T_{N-1,N}T_{N-2,N-1}\dots T_{0,1}\left(
\begin{array}
[c]{c}%
A_{0}\\
B_{0}%
\end{array}
\right) \nonumber\\
&  =\left(
\begin{array}
[c]{cc}%
T_{11} & T_{12}\\
T_{21} & T_{22}%
\end{array}
\right)  \left(
\begin{array}
[c]{c}%
A_{0}\\
B_{0}%
\end{array}
\right)  , \label{mat2}%
\end{align}
where $N$ is the total number of segments. For bound states, this equation
must be complemented by suitable boundary conditions. One possibility is to
enforce decaying solutions at the boundaries, $A_{0}=B_{N}=0$, corresponding
to $T_{22}=0$ in (\ref{mat2}), which is satisfied only for specific energies
$E$, the eigenenergies of the bound states \cite{1990IJQE...26.2025J}.

For the piecewise linear potential approach (Fig. 1(b)), the potential in each
segment $j$ is linearly interpolated, $V\left(  z\right)  =V_{j}%
+V_{z,j}\left(  z-z_{j}\right)  $ for $z_{j}\leq z\leq z_{j}+\Delta
_{j}=z_{j+1}$, with $V_{z,j}=\left(  V_{j+1}-V_{j}\right)  /\Delta_{j}$.
Equation (\ref{effm}) can then be solved analytically in terms of the Airy
functions $\mathrm{Ai}$ and $\mathrm{Bi}$ \cite{1990IJQE...26.2025J},%
\begin{equation}
\psi\left(  z\right)  =\mathcal{A}_{j}\mathrm{Ai}\left(  s_{j}+\frac{z-z_{j}%
}{\ell_{j}}\right)  +\mathcal{B}_{j}\mathrm{Bi}\left(  s_{j}+\frac{z-z_{j}%
}{\ell_{j}}\right)  \label{airy}%
\end{equation}
for $z_{j}\leq z\leq z_{j+1}$, with $s_{j}=\left(  V_{j}-E\right)
/\varepsilon_{j}$ and $\ell_{j}=\varepsilon_{j}/V_{z,j}$, where $\varepsilon
_{j}=\sqrt[3]{\hbar^{2}V_{z,j}^{2}/\left(  2m_{j}^{\ast}\right)  }$. We obtain%
\begin{align}
\psi_{j+1}  &  =\mathcal{A}_{j}\mathrm{Ai}(s_{j}+\frac{\Delta_{j}}{\ell_{j}%
})+\mathcal{B}_{j}\mathrm{Bi}(s_{j}+\frac{\Delta_{j}}{\ell_{j}}),\nonumber\\
\psi_{j+1}^{\prime}  &  =\ell_{j}^{-1}\mathcal{A}_{j}\mathrm{Ai}^{\prime
}(s_{j}+\frac{\Delta_{j}}{\ell_{j}})+\ell_{j}^{-1}\mathcal{B}_{j}%
\mathrm{Bi}^{\prime}(s_{j}+\frac{\Delta_{j}}{\ell_{j}}), \label{airy1}%
\end{align}
and%
\begin{align}
\mathcal{A}_{j}  &  =D_{j}^{-1}\mathrm{Bi}^{\prime}(s_{j})\psi_{j}-D_{j}%
^{-1}\ell_{j}\mathrm{Bi}(s_{j})\psi_{j}^{\prime},\nonumber\\
\mathcal{B}_{j}  &  =-D_{j}^{-1}\mathrm{Ai}^{\prime}(s_{j})\psi_{j}+D_{j}%
^{-1}\ell_{j}\mathrm{Ai}(s_{j})\psi_{j}^{\prime}, \label{airy2}%
\end{align}
with $D_{j}=\mathrm{Ai}(s_{j})\mathrm{Bi}^{\prime}(s_{j})-\mathrm{Ai}^{\prime
}(s_{j})\mathrm{Bi}(s_{j})$. Here a prime denotes a derivative with respect to
the argument of the Airy function (for $\mathrm{Ai}^{\prime}$, $\mathrm{Bi}%
^{\prime}$) or the position $z$ (in all other cases). A position dependent
effective mass is treated by assigning a constant value to each segment $j$,
for example $m^{\ast}\left(  z_{j}\right)  $ or preferably $\left[  m^{\ast
}\left(  z_{j}\right)  +m^{\ast}\left(  z_{j+1}\right)  \right]  /2$ (see
appendix), and using the matching conditions (\ref{match}) at the boundary
between two adjacent segments \cite{1990IJQE...26.2025J}. A piecewise linear
interpolation of $m^{\ast}$ as for the potential is not feasible, since then
the solutions of (\ref{effm}) cannot be expressed in terms of Airy functions
anymore. Equations (\ref{airy1}), (\ref{airy2}) can again be rewritten as a
matrix equation of the form (\ref{mat}), allowing us to treat the quantum
structure using (\ref{mat2}) in a similar manner as described above
\cite{1990IJQE...26.2025J}. Interfaces introducing abrupt potential changes in
the quantum structure must be taken into account explicitly in the Airy
function approach by employing the matching conditions (\ref{match}).

\subsection{Symmetrized matrix}

\label{sub:tra2}

In the transfer matrix approach, the amplitudes $A_{N}$ and $B_{N}$ at the
right boundary of the structure are related to the values $A_{0}$ and $B_{0}$
at the left boundary by repeatedly applying the transfer matrix. Due to the
segmentation of the potential, an error is introduced in (\ref{mat2}) for
every propagation step from a position $z_{j}$ to $z_{j+1}$, which is
typically characterized in terms of the local discretization error (LDE). The
LDE is defined as the difference between the exact and computed solution at a
position $z_{j+1}$ obtained from a given function value at $z_{j}$. In the
appendix, the LDE with respect to the amplitudes $A_{j}$ and $B_{j}$ for the
transfer matrix (\ref{T1}) is found to be $\mathcal{O}\left(  \Delta_{j}%
^{2}\right)  $. It can be improved to $\mathcal{O}\left(  \Delta_{j}%
^{3}\right)  $ by symmetrizing the matrix, i.e., placing the potential step in
the middle of the segment, see Fig. 1(c). The resulting transfer matrix is
then with $k_{j}^{\pm}=\left(  k_{j}\pm k_{j+1}\right)  /2$ given by%

\begin{align}
T_{j,j+1} &  =T_{j+1}\left(  \frac{\Delta_{j}}{2}\right)  T_{j\rightarrow
j+1}T_{j}\left(  \frac{\Delta_{j}}{2}\right)  \nonumber\\
&  =\left(
\begin{array}
[c]{cc}%
\frac{\beta_{j+1}+\beta_{j}}{2\beta_{j+1}}e^{\mathrm{i}k_{j}^{+}\Delta_{j}} &
\frac{\beta_{j+1}-\beta_{j}}{2\beta_{j+1}}e^{-\mathrm{i}k_{j}^{-}\Delta_{j}}\\
\frac{\beta_{j+1}-\beta_{j}}{2\beta_{j+1}}e^{\mathrm{i}k_{j}^{-}\Delta_{j}} &
\frac{\beta_{j+1}+\beta_{j}}{2\beta_{j+1}}e^{-\mathrm{i}k_{j}^{+}\Delta_{j}}%
\end{array}
\right)  ,\label{T2}%
\end{align}
where again $k_{j}=\sqrt{2m_{j}^{\ast}\left(  E-V_{j}\right)  }/\hbar$,
$\beta_{j}=k_{j}/m_{j}^{\ast}$. As in the Airy function approach, interfaces
introducing abrupt potential changes in the quantum structure must be dealt
with separately by applying the matching conditions; here, the corresponding
transfer matrix (\ref{Tst}) can be used.

\section{Comparison}

\label{sec:com}

The improved transfer matrix (\ref{T2}) can be evaluated at a comparable
computational cost as the matrix (\ref{T1}), but exhibits a superior accuracy.
As shown in the appendix, the local discretization error with respect to the
amplitudes $A_{j}$ and $B_{j}$ is improved from $\mathcal{O}\left(  \Delta
_{j}^{2}\right)  $\ to $\mathcal{O}\left(  \Delta_{j}^{3}\right)  $ for
arbitrary potentials and effective masses, i.e., the same order as for the
Airy function approach, which however involves a significantly higher
computational effort. \begin{figure}[t]
\centering\includegraphics[width=3.5in]{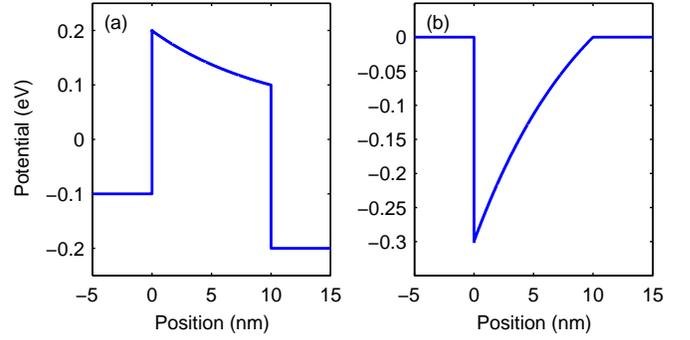}\caption{Exponential model
potential with $d=10\,\mathrm{nm}$ and $K=-1/d$, used for evaluating the
accuracy of various methods. (a) Barrier. (b) Quantum well.}%
\label{pot}%
\end{figure}

In the following, we compare the accuracy of the different methods for an
analytically solvable model potential. Here, polynomial test potentials are
not suitable for a general discussion since their higher order derivatives
identically vanish, which can lead to an increased accuracy in such special
cases. Especially triangular or other piecewise linear potentials are
obviously inadequate since the Airy function approach then becomes exact.
Instead, we choose the exponential ansatz
\begin{equation}
V(z)=V_{0}+V_{1}\exp\left(  Kz\right)  , \label{exppot}%
\end{equation}
$0\leq z\leq d$ (see Fig. \ref{pot}), approaching a linear function for
$K\rightarrow0$. Such a potential can for example serve as a model for the
effective potential profile in the presence of space charges
\cite{2002IEDL...23..348S,2005JAP....97f4107N}.

\subsection{Position independent effective mass}

For now, we assume a constant effective mass $m^{\ast}$. Then, analytical
solutions of the form
\begin{equation}
\psi=c_{1}\mathrm{J}_{\mu}\left(  a\right)  +c_{2}\mathrm{Y}_{\mu}\left(
a\right)  \label{bess}%
\end{equation}
exist for the potential (\ref{exppot}), with constants $c_{1}$ and $c_{2}$.
Here, $\mathrm{J}_{\mu}$\ and $\mathrm{Y}_{\mu}$ are Bessel functions of the
first and second kind,\ and the parameters are given by%
\begin{align}
\mu &  =2\frac{\sqrt{2m^{\ast}\left(  V_{0}-E\right)  }}{\hbar K},\nonumber\\
a(z)  &  =2\frac{\sqrt{-2m^{\ast}V_{1}}}{\hbar K}\exp\left(  \frac{1}%
{2}Kz\right)  .
\end{align}

\begin{figure}[t]
\centering\includegraphics[width=3.5in]{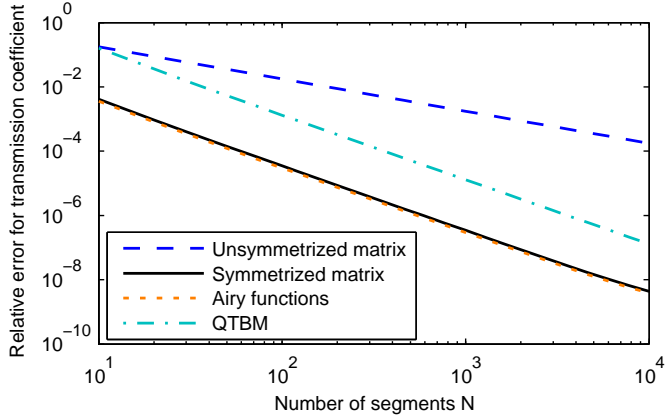}\caption{Relative error
$\varepsilon_{T}=\left|  1-T_{\mathrm{num}}/T\right|  $ of the numerically
obtained transmission coefficient $T_{\mathrm{num}}$ as a function of the
number of segments $N$. The corresponding barrier is shown in Fig.
\ref{pot}(a), the effective mass is assumed to be constant.}%
\label{Tran}%
\end{figure}

For our simulations, the different transfer matrix approaches discussed in
Section \ref{sec:tra} are used to compute an overall matrix based on
(\ref{mat2}), from which the required quantities can be extracted. First, we
investigate the barrier structure shown in Fig. \ref{pot}(a), which can be
characterized in terms of a transmission coefficient $T$, giving the tunneling
probability of an electron \cite{1997plds.book.....D}. The unsymmetrized,
symmetrized and Airy function transfer matrix approaches are evaluated, based
on the expressions (\ref{T1}), (\ref{T2}) and (\ref{airy1}), respectively; for
comparison, also the QTBM result\ is computed. Assuming an electron energy of
$E=0$ and a constant effective mass of $m^{\ast}=0.067\,m_{e}$ corresponding
to GaAs, where $m_{e}$ is the electron mass, the exact value obtained by
evaluating (\ref{bess}) is $T=1.749\times10^{-4}$. Fig. \ref{Tran} shows the
relative error $\varepsilon_{T}\left(  N\right)  =\left|  1-T_{\mathrm{num}%
}\left(  N\right)  /T\right|  $ as a function of $N\propto\Delta^{-1}$. Here,
$T_{\mathrm{num}}\left(  N\right)  $ is the numerical result for the
transmission coefficient, as obtained by the different methods for a
subdivision of the structure into $N$ segments of equal length $\Delta
=d/N\propto N^{-1}$. As can be seen from Fig. \ref{Tran}, the error scales
with $N^{-1}\propto\Delta$ for the unsymmetrized transfer matrix approach and
with $N^{-2}\propto\Delta^{2}$ for the other methods. This can easily be
understood by means of the local discretization error, which is $\mathcal{O}%
\left(  \Delta^{3}\right)  $ for the Airy function approach and the
symmetrized transfer matrix, and $\mathcal{O}\left(  \Delta^{2}\right)  $ for
the unsymmetrized matrix, as discussed above and in the appendix. When the
overall transfer matrix of the structure is computed from (\ref{mat2}), the
individual LDEs arising for each of the $N$ segments accumulate, thus
resulting in a total error $N\mathcal{O}\left(  \Delta^{2}\right)
=\mathcal{O}\left(  \Delta\right)  $ for the unsymmetrized approach and
$N\mathcal{O}\left(  \Delta^{3}\right)  =\mathcal{O}\left(  \Delta^{2}\right)
$\ for the other schemes.

The symmetrized transfer matrix approach and the Airy function method are the
most accurate, both exhibiting a comparable error $\varepsilon_{T}\left(
N\right)  $. However, the symmetrized matrix approach is much more
computationally efficient, being over $20$ times faster than the Airy function
method in our MATLAB implementation. For a given $N$, the QTBM is even three
times faster than the symmetrized matrix approach, but also 40 times less
accurate, meaning that it requires $\sqrt{40}\approx6$ times as many grid
points as the symmetrized matrix approach to achieve the same
accuracy.\begin{figure}[t]
\centering\includegraphics[width=3.5in]{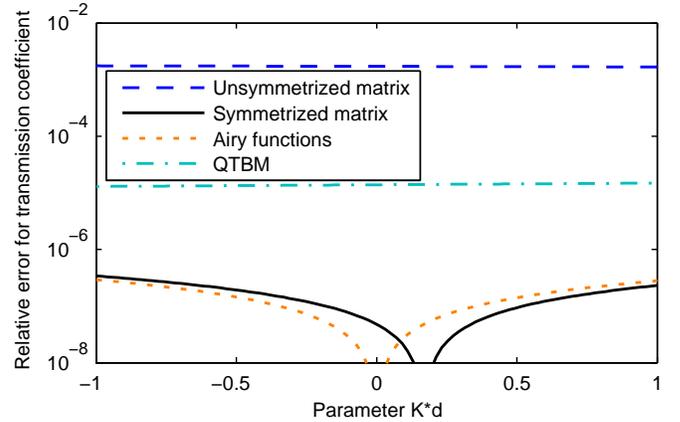}\caption{Relative error
$\varepsilon_{T}=\left|  1-T_{\mathrm{num}}/T\right|  $ of the numerically
obtained transmission coefficient $T_{\mathrm{num}}$ as a function of $Kd$.
Here, $N=1000$ segments are used. The corresponding barrier is shown in Fig.
\ref{pot}(a), the effective mass is assumed to be constant.}%
\label{Kd}%
\end{figure}

Fig. (\ref{Kd}) shows again the relative error $\varepsilon_{T}$, but now for
a fixed number of segments $N=1000$. Instead, the shape of the potential is
modified by varying $K$ in (\ref{exppot}), and also adapting $V_{0}$ and
$V_{1}$ so that $V\left(  z\right)  $ remains constant at $z=0$ and $z=d$ and
only the curvature of the potential changes. The symmetrized matrix approach
and the Airy function method exhibit a superior accuracy especially for small
$K$, corresponding to a weak curvature of the potential. While the error of
the unsymmetrized matrix approach and the QTBM show only a weak dependence on
$K$, the Airy function method becomes exact for $K\rightarrow0$, where the
potential becomes piecewise linear. Interestingly, also the symmetrized
transfer matrix approach has a vanishing error $\varepsilon_{T}$ for a
specific value of $K$, at $Kd\approx0.167$.

\begin{figure}[t]
\centering\includegraphics[width=3.5in]{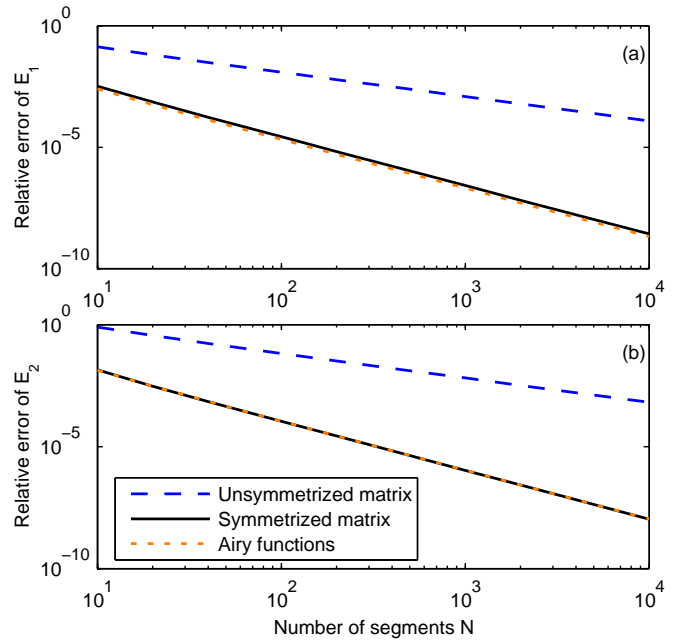}\caption{Relative error
$\varepsilon_{E}=\left|  1-E_{\mathrm{num}}/E\right|  $ of the numerically
obtained eigenenergy $E_{\mathrm{num}}$ for the (a)\ first and (b) second
bound state as a function of the number of segments $N$. The corresponding
well is shown in Fig. \ref{pot}(b), the effective mass is assumed to be
constant.}%
\label{E}%
\end{figure}

Now we apply the different numerical methods to the bound states of the
potential well shown in Fig. \ref{pot}(b). Again assuming a constant effective
mass of $m^{\ast}=0.067\,m_{e}$, evaluation of (\ref{bess}) yields two bound
states with eigenvalues $E_{1}=-0.1343$\thinspace$\mathrm{eV}$ and
$E_{2}=-0.0129\,\mathrm{eV}$, respectively. In the following, we compare the
accuracy of the numerically found eigenenergies $E_{\mathrm{num}}$, as
obtained by the unsymmetrized and the symmetrized transfer matrix approach and
the Airy function method, corresponding to the expressions (\ref{T1}),
(\ref{T2}) and (\ref{airy1}), respectively. Here, we again divide the
structure into $N$ segments of equal length $\Delta=d/N\propto N^{-1}$. Fig.
\ref{E} shows the relative error $\varepsilon_{E}\left(  N\right)  =\left|
1-E_{\mathrm{num}}\left(  N\right)  /E\right|  $ for the first and the second
bound state as a function of $N$. As for the transmission coefficient in Fig.
\ref{Tran}, the error scales with $N^{-1}\propto\Delta$ for the unsymmetrized
matrix approach and with $N^{-2}\propto\Delta^{2}$ for the other methods.
Again, the symmetrized matrix approach and the Airy function method exhibit a
comparable value of $\varepsilon_{E}\left(  N\right)  $, being far superior to
the unsymmetrized matrix approach.

\subsection{Position dependent effective mass}

Now we compare the accuracy of the different methods for a position dependent
effective mass $m^{\ast}(z)$. Here, we choose the same exponential ansatz for
the potential as above, see (\ref{exppot}) and Fig. \ref{pot}. For an
effective mass of the form $m^{\ast}=m_{0}^{\ast}\exp\left(  -Kz\right)  $,
again\ an analytical solution exists:
\begin{equation}
\psi=c_{1}\mathrm{J}_{\mu}\left(  a\right)  \exp\left(  -Kz/2\right)
+c_{2}\mathrm{Y}_{\mu}\left(  a\right)  \exp\left(  -Kz/2\right)  ,
\label{bess2}%
\end{equation}
with%
\begin{align}
\mu &  =-\sqrt{1+8\frac{m_{0}V_{1}}{K^{2}\hbar^{2}}},\nonumber\\
a(z)  &  =2\frac{\sqrt{-2m_{0}\left(  V_{0}-E\right)  }}{\hbar K}\exp\left(
-\frac{1}{2}Kz\right)  .
\end{align}
The transfer matrix definitions (\ref{T1}), (\ref{T2}) are also valid for
position dependent effective masses. In the Airy function approach
(\ref{airy1}), a position dependent effective mass can be accounted for by
assuming a constant value within each segment, as discussed at the end of
Section \ref{sub:tra1}. Here, we assign the averaged mass $\left(  m_{j}%
^{\ast}+m_{j+1}^{\ast}\right)  /2$ rather than $m_{j}^{\ast}$\ to each
segment, since then the third order LDE, found for the amplitudes
$\mathcal{A}$ and $\mathcal{B}$ in the case of position independent masses, is
also preserved for the position dependent case, see the
appendix.\begin{figure}[t]
\centering\includegraphics[width=3.5in]{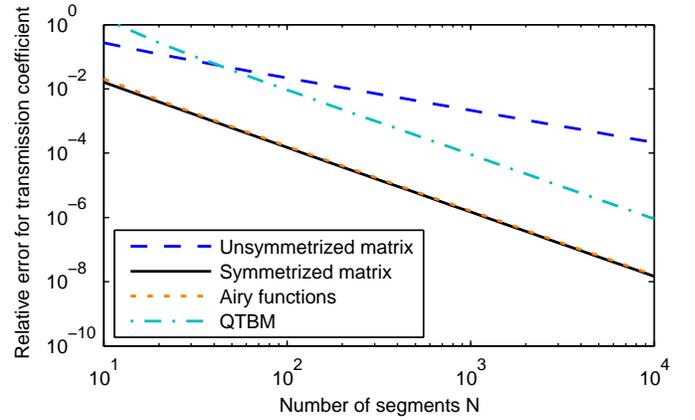}\caption{Relative error
$\varepsilon_{T}=\left|  1-T_{\mathrm{num}}/T\right|  $ of the numerically
obtained transmission coefficient $T_{\mathrm{num}}$ as a function of the
number of segments $N$. The corresponding barrier is shown in Fig.
\ref{pot}(a), the effective mass is assumed to be position dependent.}%
\label{Tranm}%
\end{figure}

Fig. \ref{Tranm} corresponds to Fig. \ref{Tran}, but now for a position
dependent effective potential with $m^{\ast}=0.2\,m_{e}\exp\left(  -Kz\right)
$ for $0\leq z\leq d$ and $m^{\ast}=0.067\,m_{e}$ otherwise. The exact
transmission coefficient for an electron with energy $E=0$, as obtained by
evaluating (\ref{bess2}), is now $T=5.376\times10^{-10}$. From Fig.
\ref{Tranm} we can see that also here the error scales with $N^{-1}%
\propto\Delta$ for the unsymmetrized matrix approach and with $N^{-2}%
\propto\Delta^{2}$ for the other methods, compare Fig. \ref{Tran}. Again, the
symmetrized matrix approach and the Airy function method are the most
accurate, with the symmetrized matrix approach being numerically much more
efficient. \begin{figure}[t]
\centering\includegraphics[width=3.5in]{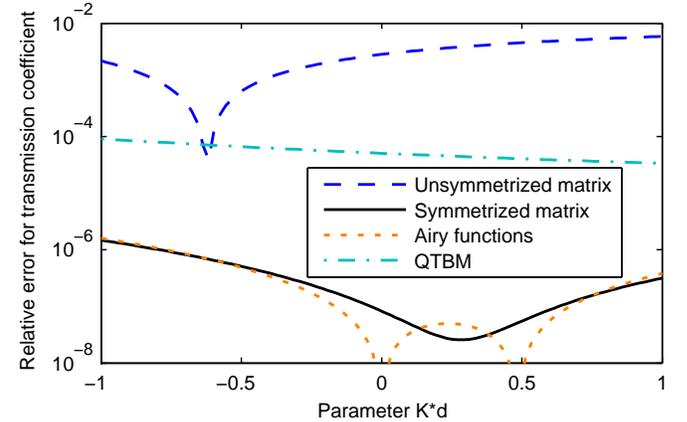}\caption{Relative error
$\varepsilon_{T}=\left|  1-T_{\mathrm{num}}/T\right|  $ of the numerically
obtained transmission coefficient $T_{\mathrm{num}}$ as a function of $Kd$.
Here, $N=1000$ segments are used. The corresponding barrier is shown in Fig.
\ref{pot}(a), the effective mass is assumed to be position dependent.}%
\label{Kd_m}%
\end{figure}

For the sake of completeness, Fig. (\ref{Kd_m}) is shown as the counterpart of
Fig. \ref{Kd}, but now taking into account a position dependent effective mass
as above. Again, the symmetrized matrix approach and the Airy function method
have a superior accuracy especially for small values of $K$, corresponding to
a weak curvature of the potential.

\section{Example: Schr\"{o}dinger-Poisson solver}

\label{sec:ex}

In the following, we apply the transfer matrices discussed above to a
real-world example, namely finding the wavefunctions and eigenenergies of the
quantum cascade laser (QCL) structure described in \cite{2001ApPhL..78.3529P}.
The goal is to evaluate and compare the performance of the different
approaches for a practical problem, and to discuss the inclusion of additional
important effects. Specifically, we here also account for energy-band
nonparabolicity, and complement the Schr\"{o}dinger equation by the Poisson
equation to take into account space charge effects. In practice, extensive
parameter scans have to be performed for QCL design optimization. Thus, the
simulation of QCLs calls for especially efficient methods, the more so as the
self-consistent solution of the Schr\"{o}dinger-Poisson system results in a
further increase of the numerical effort.

In simulations, the QCL structure is defined by an infinitely repeated
elementary sequence of multiple wells and barriers (called a period). For such
a structure under bias, it is sufficient to compute the eigenenergies and
corresponding wave functions for a single energy interval given by the bias
across one period; the solutions of the other periods are then obtained by
appropriate shifts in position and energy. We solve the Schr\"{o}dinger
equation using the approaches defined by (\ref{T1}), (\ref{T2}) and
(\ref{airy1}), respectively. For all three methods, we treat band edge
discontinuities at the barrier-well interfaces explicitly using the matching
conditions (\ref{match}), to obtain an optimum accuracy. We use a simulation
window of four periods to keep the influence of the boundaries negligible, and
determine the bound states similarly as in Section \ref{sec:com}. To combine
reasonable numerical efficiency with a good accuracy, we choose a segment
length of $\Delta=2$\thinspace$\mathrm{nm}$ (the last segment of each well or
barrier is $\Delta\leq2$\thinspace$\mathrm{nm}$).

Various models are available for including nonparabolicity
\cite{1987PhRvB..35.7770N}; here, we use an energy dependent effective mass
$m_{j}^{\ast}\left(  E\right)  =m_{j}^{\ast}\left[  1+\left(  E-V_{j}\right)
/E_{g,j}\right]  $ (with band gap energy $E_{g,j}$ at position $z_{j}$), which
can straightforwardly be implemented into the transfer matrices. The Poisson
equation is given by \cite{2000JAP....87.7931C,2008SeScT..23l5040L}%
\begin{equation}
-\partial_{z}\left[  \epsilon\left(  z\right)  \partial_{z}\varphi\left(  z\right)
\right]  =e\left[  N\left(  z\right)  -\sum_{n}n_{\mathrm{2D},n}\left|
\psi_{n}\left(  z\right)  \right|  ^{2}\right]  ,\label{poisson}%
\end{equation}
leading to an additional potential -e$\varphi$ in (\ref{effm}). Here, $\epsilon\left(  z\right)  $ is the permittivity, $e$ is the elementary
charge, $N\left(  z\right)  $ is the doping concentration, and $n_{\mathrm{2D}%
,n}$ is the electron sheet density of level $n$ with wave function $\psi
_{n}\left(  z\right)  $. While for an operating QCL, $n_{\mathrm{2D},n}$ can
only be exactly determined by detailed carrier transport simulations
\cite{2007JAP...101h6109J}, this is prohibitive for design optimizations of
experimental QCL structures over an extended parameter range. Thus, for
solving the Schr\"{o}dinger-Poisson system, simpler and much faster approaches
are commonly adopted, such as applying Fermi-Dirac statistics
\cite{2000JAP....87.7931C,2008SeScT..23l5040L}
\begin{equation}
n_{\mathrm{2D},n}=\frac{m^{\ast}}{\pi\hbar^{2}}k_{\mathrm{B}}T\ln\left(
1+\exp\left[  \left(  \mu-\tilde{E}_{n}\right)  /\left(  k_{\mathrm{B}%
}T\right)  \right]  \right)  ,\label{fermi}%
\end{equation}
where $\mu$ is the chemical potential, $k_{\mathrm{B}}$ is the Boltzmann
constant, $T$ is the lattice temperature, and $m^{\ast}$ is the effective
mass, here taken to be the value of the well material. In (\ref{fermi}), we
use the energy of a state relative to the conduction band edge $\tilde{E}%
_{n}=E_{n}-\int V\left|  \psi_{n}\right|  ^{2}\mathrm{d}z$ rather than $E_{n}$
itself to correctly reflect the invariance properties of the biased structure.
Especially, this ensures that the simulation results do not depend on the
choice of the elementary period in the structure. The chemical potential $\mu$
is found from the charge neutrality condition within one period. The
Schr\"{o}dinger and Poisson equations are iteratively solved until
self-consistency is achieved. For the Poisson equation (\ref{poisson}), we
employ a finite difference scheme on a $1$\thinspace\textrm{\AA}-grid, where
we use (\ref{psi}) and (\ref{airy}) to appropriately interpolate the
eigenfunctions obtained from the Schr\"{o}dinger solver. \begin{figure}[t]
\centering\includegraphics[width=3.5in]{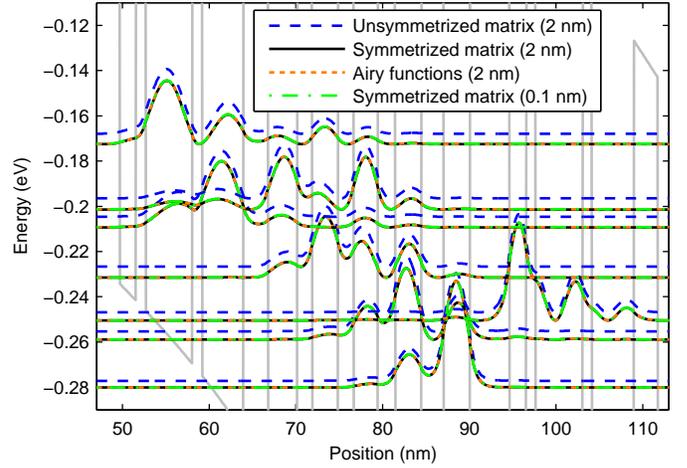}\caption{Self-consistent
band profile (grey line), energy levels and wave functions squared for the
QCL\ in \cite{2001ApPhL..78.3529P} at a bias of $48$\thinspace$\mathrm{kV/cm}$
and a temperature of $300$\thinspace$\mathrm{K}$. Shown are the results as
obtained with the three transfer matrix methods for $\Delta=2$\thinspace
$\mathrm{nm}$, and also the symmetrized matrix result for $\Delta
=0.1$\thinspace$\mathrm{nm}$, which practically coincides with the symmetrized
matrix and Airy function results for $\Delta=2$\thinspace$\mathrm{nm}$.}%
\label{QCL}%
\end{figure}

Simulations of the QCL in \cite{2001ApPhL..78.3529P} have been performed at
various temperatures, considering the seven lowest levels (i.e, with lowest
energies $\tilde{E}_{n}$) within each period. In Fig. (\ref{QCL}), the
obtained energy levels and wave functions squared of a single period are shown
for the unsymmetrized, symmetrized and Airy function matrix approach at
$T=300$\thinspace$\mathrm{K}$, using a segment length of $\Delta=2$%
\thinspace$\mathrm{nm}$. Also the symmetrized transfer matrix result for
$\Delta=0.1$\thinspace$\mathrm{nm}$ is plotted for reference. The symmetrized
matrix and Airy function results exhibit a similar accuracy, with deviations
in eigenenergies of around $0.1$\thinspace$\mathrm{meV}$ from the
high-accuracy result obtained with $\Delta=0.1$\thinspace$\mathrm{nm}$. In
Fig. (\ref{QCL}), those three curves are practically indistinguishable. On the
other hand, the unsymmetrized matrix method produces deviations of around
$5$\thinspace$\mathrm{meV}$. The unsymmetrized and symmetrized matrix approach
require approximately the same computation time for obtaining the
self-consistent result in Fig. (\ref{QCL}), while the Schr\"{o}dinger-Poisson
solver based on the Airy functions is about $10$ times slower. This confirms
that the symmetrized transfer matrix method combines high numerical efficiency
with excellent accuracy for practical applications.

\section{Conclusion}

In conclusion, we have compared the accuracy of different transfer matrix
approaches, as used for solving the effective mass Schr\"{o}dinger equation
with an arbitrary one-dimensional potential and a constant or position
dependent effective mass. In particular, the local discretization error has
been derived for the Airy function approach resulting from a piecewise linear
approximation of the potential, and for unsymmetrized and symmetrized transfer
matrices based on a piecewise constant potential approximation. Furthermore,
numerical simulations have been performed to evaluate the numerical accuracy
of the different approaches for scattering and bound states, employing
exponential test potentials. Comparisons to the finite difference method,
specifically the QTBM, have also been carried out. Additionally,
self-consistent Schr\"{o}dinger-Poisson device simulations are presented.

The symmetrized transfer matrix approach and the Airy function method exhibit
a comparable accuracy, being superior to the other methods investigated.
However, the symmetrized matrix approach achieves this at a significantly
reduced numerical cost, moreover avoiding the numerical problems associated
with Airy functions. All in all, the symmetrized transfer matrix approach is
shown to combine the numerical efficiency and straightforwardness of its
unsymmetrized counterpart with the superior accuracy of the Airy function method.

\appendices

\section{Local discretization error}

In the following, we derive the local discretization error (LDE) for the
different types of transfer matrices. In this context, we investigate the
piecewise constant potential approximation based on matrix (\ref{T1}) and its
symmetrized version (\ref{T2}), as well as the piecewise linear potential
scheme (\ref{airy1}). As mentioned in Section \ref{sec:tra}, the segments are
chosen so that band edge discontinuities in the structure coincide with the
borders between two segments, enabling an exact treatment in terms of the
matching conditions (\ref{match}). Thus, for our error analysis we imply that
the potential and effective mass vary smoothly within each segment, i.e., have
a sufficient degree of differentiability. Otherwise, no further assumptions
about the potential shape\ and effective mass are made. The local
discretization error for $\psi$ at $z=z_{j+1}$ is%
\begin{equation}
\tau_{j+1}^{\psi}=\psi_{j+1}-\psi\left(  z_{j+1}\right)  , \label{tau}%
\end{equation}
where $\psi_{j+1}$ is the approximate wavefunction value at $z_{j+1}$ obtained
by the transfer matrix approach from a given value $\psi\left(  z_{j}\right)
=\psi_{j}$ at $z_{j}$, while $\psi\left(  z_{j+1}\right)  $ is the exact
solution. For evaluating the LDE, it is helpful to express $\psi\left(
z_{j+1}\right)  $ in terms of a Taylor series,
\begin{align}
\psi\left(  z_{j+1}\right)   &  =\psi_{j}+\psi_{j}^{\prime}\Delta_{j}+\frac
{1}{2}\psi_{j}^{\prime\prime}\Delta_{j}^{2}+\frac{1}{6}\psi_{j}^{\left(
3\right)  }\Delta_{j}^{3}\nonumber\\
&  +\frac{1}{24}\psi_{j}^{\left(  4\right)  }\Delta_{j}^{4}+\mathcal{O}\left(
\Delta_{j}^{5}\right)  . \label{taylor}%
\end{align}
Analogously, we can define an LDE for the derivative $\psi^{\prime}$,%
\begin{equation}
\tau_{j+1}^{\psi^{\prime}}=\psi_{j+1}^{\prime}-\psi^{\prime}\left(
z_{j+1}\right)  , \label{taus}%
\end{equation}
and express $\psi^{\prime}\left(  z_{j+1}\right)  $ as%
\begin{equation}
\psi^{\prime}\left(  z_{j+1}\right)  =\psi_{j}^{\prime}+\psi_{j}^{\prime
\prime}\Delta_{j}+\frac{1}{2}\psi_{j}^{\left(  3\right)  }\Delta_{j}^{2}%
+\frac{1}{6}\psi_{j}^{\left(  4\right)  }\Delta_{j}^{3}+\mathcal{O}\left(
\Delta_{j}^{4}\right)  . \label{taylors}%
\end{equation}

\subsection{Piecewise constant potential approximation}

Using (\ref{psi}), we can relate $A_{j}$ and $B_{j}$ to the wavefunction at
position $z_{j}$,
\begin{align}
A_{j}  &  =\frac{1}{2}\left(  \psi_{j}-\mathrm{i}\frac{1}{k_{j}}\psi
_{j}^{\prime}\right)  ,\nonumber\\
B_{j}  &  =\frac{1}{2}\left(  \psi_{j}+\mathrm{i}\frac{1}{k_{j}}\psi
_{j}^{\prime}\right)  , \label{AB}%
\end{align}
and express the LDEs for the amplitudes $A$ and $B$ in terms of $\tau
_{j+1}^{\psi}$\ and $\tau_{j+1}^{\psi^{\prime}}$,%
\begin{align}
\tau_{j+1}^{A}  &  =A_{j+1}-A\left(  z_{j+1}\right)  =\frac{1}{2}\left(
\tau_{j+1}^{\psi}-\mathrm{i}\frac{1}{k_{j+1}}\tau_{j+1}^{\psi^{\prime}%
}\right)  ,\nonumber\\
\tau_{j+1}^{B}  &  =B_{j+1}-B\left(  z_{j+1}\right)  =\frac{1}{2}\left(
\tau_{j+1}^{\psi}+\mathrm{i}\frac{1}{k_{j+1}}\tau_{j+1}^{\psi^{\prime}%
}\right)  . \label{tauAB}%
\end{align}
For the unsymmetrized transfer matrix, we obtain from (\ref{psi}) with the
expressions (\ref{mat}) and (\ref{T1})%
\begin{align}
\psi_{j+1}  &  =A_{j+1}+B_{j+1}=\exp\left(  \mathrm{i}k_{j}\Delta_{j}\right)
A_{j}+\exp\left(  -\mathrm{i}k_{j}\Delta_{j}\right)  B_{j},\nonumber\\
\psi_{j+1}^{\prime}  &  =\mathrm{i}k_{j+1}\left(  A_{j+1}-B_{j+1}\right)
\nonumber\\
&  =\mathrm{i}k_{j+1}\frac{\beta_{j}}{\beta_{j+1}}\left[  A_{j}\exp\left(
\mathrm{i}k_{j}\Delta_{j}\right)  -B_{j}\exp\left(  -\mathrm{i}k_{j}\Delta
_{j}\right)  \right]  . \label{psip}%
\end{align}
For calculating the LDE, we insert the expressions (\ref{psip}) and
(\ref{taylor}) into (\ref{tau}), where we express $A_{j}$ and $B_{j}$ by
(\ref{AB}) and use (\ref{effm}) to rewrite the derivatives $\psi_{j}^{\left(
n\right)  }$ in (\ref{taylor}) as
\begin{align}
\psi_{j}^{\prime\prime}  &  =-k_{j}^{2}\psi_{j}+\frac{m_{j}^{\ast\prime}%
}{m_{j}^{\ast}}\psi_{j}^{\prime},\nonumber\\
\psi_{j}^{\left(  3\right)  }  &  =-\frac{\left(  m_{j}^{\ast}k_{j}%
^{2}\right)  ^{\prime}}{m_{j}^{\ast}}\psi_{j}-k_{j}^{2}\psi_{j}^{\prime}%
+\frac{m_{j}^{\ast\prime\prime}}{m_{j}^{\ast}}\psi_{j}^{\prime},
\end{align}
with $k_{j}=\sqrt{2m_{j}^{\ast}\left(  E-V_{j}\right)  }/\hbar$. A Taylor
expansion then yields%
\begin{align}
\tau_{j+1}^{\psi}  &  =\cos\left(  k_{j}\Delta_{j}\right)  \psi_{j}%
+\sin\left(  k_{j}\Delta_{j}\right)  k_{j}^{-1}\psi_{j}^{\prime}\nonumber\\
&  -\psi_{j}-\psi_{j}^{\prime}\Delta_{j}-\frac{1}{2}\psi_{j}^{\prime\prime
}\Delta_{j}^{2}-\frac{1}{6}\psi_{j}^{\left(  3\right)  }\Delta_{j}%
^{3}+\mathcal{O}\left(  \Delta_{j}^{4}\right) \nonumber\\
&  =-\frac{1}{2}\frac{m_{j}^{\ast\prime}}{m_{j}^{\ast}}\psi_{j}^{\prime}%
\Delta_{j}^{2}+\frac{1}{6}\left(  \frac{\left(  m_{j}^{\ast}k_{j}^{2}\right)
^{\prime}}{m_{j}^{\ast}}\psi_{j}-\frac{m_{j}^{\ast\prime\prime}}{m_{j}^{\ast}%
}\psi_{j}^{\prime}\right)  \Delta_{j}^{3}\nonumber\\
&  +\mathcal{O}\left(  \Delta_{j}^{4}\right)  .
\end{align}
Analogously, by inserting the expressions (\ref{psip}) and (\ref{taylors})
into (\ref{taus}) we obtain%
\begin{align}
\tau_{j+1}^{\psi^{\prime}}  &  =\frac{m_{j+1}^{\ast}}{m_{j}^{\ast}}\left(
\psi_{j}^{\prime}\cos\left(  k_{j}\Delta_{j}\right)  -k_{j}\sin\left(
k_{j}\Delta_{j}\right)  \psi_{j}\right) \nonumber\\
&  -\psi_{j}^{\prime}-\psi_{j}^{\prime\prime}\Delta_{j}-\frac{1}{2}\psi
_{j}^{\left(  3\right)  }\Delta_{j}^{2}+\mathcal{O}\left(  \Delta_{j}%
^{3}\right) \nonumber\\
&  =\left(  \frac{m_{j+1}^{\ast}}{m_{j}^{\ast}}-1-\frac{m_{j}^{\ast\prime}%
}{m_{j}^{\ast}}\Delta_{j}\right)  \psi_{j}^{\prime}+\frac{m_{j}^{\ast}%
-m_{j+1}^{\ast}}{m_{j}^{\ast}}k_{j}^{2}\psi_{j}\Delta_{j}\nonumber\\
&  +\frac{1}{2}\left(  \frac{\left(  m_{j}^{\ast}k_{j}^{2}\right)  ^{\prime}%
}{m_{j}^{\ast}}\psi_{j}-\frac{m_{j}^{\ast\prime\prime}}{m_{j}^{\ast}}\psi
_{j}^{\prime}+\frac{m_{j}^{\ast}-m_{j+1}^{\ast}}{m_{j}^{\ast}}k_{j}^{2}%
\psi_{j}^{\prime}\right)  \Delta_{j}^{2}\nonumber\\
&  +\mathcal{O}\left(  \Delta_{j}^{3}\right) \nonumber\\
&  =\left(  \frac{\left(  m_{j}^{\ast}k_{j}^{2}\right)  ^{\prime}}%
{2m_{j}^{\ast}}-\frac{m_{j}^{\ast\prime}}{m_{j}^{\ast}}k_{j}^{2}\right)
\psi_{j}\Delta_{j}^{2}+\mathcal{O}\left(  \Delta_{j}^{3}\right)  ,
\label{tau1s}%
\end{align}
where we use $m_{j+1}^{\ast}=m_{j}^{\ast}+m_{j}^{\ast\prime}\Delta_{j}%
+\frac{1}{2}m_{j}^{\ast\prime\prime}\Delta_{j}^{2}+\mathcal{O}\left(
\Delta_{j}^{3}\right)  $ to obtain the last line of (\ref{tau1s}). Thus,
$\tau_{j+1}^{\psi}\ $is $\mathcal{O}\left(  \Delta_{j}^{2}\right)  $
($\mathcal{O}\left(  \Delta_{j}^{3}\right)  $\ for a constant effective mass),
and $\tau_{j+1}^{\psi^{\prime}}=\mathcal{O}\left(  \Delta_{j}^{2}\right)  $.
With (\ref{tauAB}), we see that both $\tau_{j+1}^{A}$ and $\tau_{j+1}^{B}$ are
$\mathcal{O}\left(  \Delta_{j}^{2}\right)  $.

In a similar manner, we obtain for the symmetrized matrix (\ref{T2})
$\tau_{j+1}^{\psi}=\mathcal{O}\left(  \Delta_{j}^{3}\right)  $ and $\tau
_{j+1}^{\psi^{\prime}}=\mathcal{O}\left(  \Delta_{j}^{3}\right)  $. More
precisely, the calculation yields for a constant $m^{\ast}$
\begin{align}
\tau_{j+1}^{\psi}  &  =\frac{1}{24}\left(  k_{j}^{2}\right)  ^{\prime}\psi
_{j}\Delta_{j}^{3}+\mathcal{O}\left(  \Delta_{j}^{4}\right)  ,\nonumber\\
\tau_{j+1}^{\psi^{\prime}}  &  =-\frac{1}{12}\left(  k_{j}^{2}\right)
^{\prime\prime}\psi_{j}\Delta_{j}^{3}-\frac{1}{24}\left(  k_{j}^{2}\right)
^{\prime}\psi_{j}^{\prime}\Delta_{j}^{3}+\mathcal{O}\left(  \Delta_{j}%
^{4}\right)
\end{align}
(and a somewhat more complicated expression for a position dependent effective
mass). This means that $\tau_{j+1}^{A}$ and $\tau_{j+1}^{B}$ are now
$\mathcal{O}\left(  \Delta_{j}^{3}\right)  $.

\subsection{Piecewise linear potential approximation}

For computing the LDEs (\ref{tau}) and (\ref{taus}) of the Airy function
approach, we proceed in a manner similar as above. Equations (\ref{airy1}) and
(\ref{airy2}) give the relation between values $\psi_{j}$, $\psi_{j}^{\prime}$
at $z_{j}$ and the numerical result $\psi_{j+1}$, $\psi_{j+1}^{\prime}$
obtained at $z_{j+1}$ from the Airy function approach:%

\begin{align}
\psi_{j+1}  &  =\frac{1}{D_{j}}\left[  \mathrm{Ai}(s_{j}+\frac{\Delta_{j}%
}{\ell_{j}})\mathrm{Bi}^{\prime}(s_{j})-\mathrm{Ai}^{\prime}(s_{j}%
)\mathrm{Bi}(s_{j}+\frac{\Delta_{j}}{\ell_{j}})\right]  \psi_{j}\nonumber\\
&  +\frac{\ell_{j}}{D_{j}}\left[  \mathrm{Ai}(s_{j})\mathrm{Bi}(s_{j}%
+\frac{\Delta_{j}}{\ell_{j}})-\mathrm{Ai}(s_{j}+\frac{\Delta_{j}}{\ell_{j}%
})\mathrm{Bi}(s_{j})\right]  \psi_{j}^{\prime}\nonumber\\
&  =\left(  1+\frac{1}{2}\frac{s_{j}}{\ell_{j}^{2}}\Delta_{j}^{2}+\frac{1}%
{6}\frac{\Delta_{j}^{3}}{\ell_{j}^{3}}+\frac{1}{24}\frac{s_{j}^{2}}{\ell
_{j}^{4}}\Delta_{j}^{4}\right)  \psi_{j}\nonumber\\
&  +\left(  \Delta_{j}+\frac{1}{6}\frac{s_{j}}{\ell_{j}^{2}}\Delta_{j}%
^{3}+\frac{1}{12}\frac{\Delta_{j}^{4}}{\ell_{j}^{3}}\right)  \psi_{j}^{\prime
}+\mathcal{O}\left(  \Delta_{j}^{5}\right)  , \label{airy3}%
\end{align}%
\begin{align}
\psi_{j+1}^{\prime}  &  =\frac{1}{D_{j}\ell_{j}}\left[  \mathrm{Ai}^{\prime
}(s_{j}+\frac{\Delta_{j}}{\ell_{j}})\mathrm{Bi}^{\prime}(s_{j})\right.
\nonumber\\
&  \left.  -\mathrm{Bi}^{\prime}(s_{j}+\frac{\Delta_{j}}{\ell_{j}}%
)\mathrm{Ai}^{\prime}(s_{j})\right]  \psi_{j}\nonumber\\
&  +\frac{1}{D_{j}}\left[  \mathrm{Ai}(s_{j})\mathrm{Bi}^{\prime}(s_{j}%
+\frac{\Delta_{j}}{\ell_{j}})-\mathrm{Ai}^{\prime}(s_{j}+\frac{\Delta_{j}%
}{\ell_{j}})\mathrm{Bi}(s_{j})\right]  \psi_{j}^{\prime}\nonumber\\
&  =\left(  s_{j}\frac{\Delta_{j}}{\ell_{j}^{2}}+\frac{1}{2}\frac{1}{\ell
_{j}^{3}}\Delta_{j}^{2}+\frac{1}{6}\frac{s_{j}^{2}}{\ell_{j}^{4}}\Delta
_{j}^{3}\right)  \psi_{j}\nonumber\\
&  +\left(  1+\frac{1}{2}\frac{s_{j}}{\ell_{j}^{2}}\Delta_{j}^{2}+\frac{1}%
{3}\frac{\Delta_{j}^{3}}{\ell_{j}^{3}}\right)  \psi_{j}^{\prime}%
+\mathcal{O}\left(  \Delta_{j}^{4}\right)  , \label{airy4}%
\end{align}
with $D_{j}=\mathrm{Ai}(s_{j})\mathrm{Bi}^{\prime}(s_{j})-\mathrm{Ai}^{\prime
}(s_{j})\mathrm{Bi}(s_{j})$. The exact results $\psi\left(  z_{j+1}\right)  $
and $\psi^{\prime}\left(  z_{j+1}\right)  $ are again expressed by the Taylor
series expansions (\ref{taylor}) and (\ref{taylors}), respectively, where we
rewrite the derivatives $\psi_{j}^{\left(  n\right)  }$ in terms of $\psi_{j}$
and $\psi_{j}^{\prime}$. For a constant effective mass, we have
\begin{align}
\psi_{j}^{\prime\prime}  &  =\ell_{j}^{-2}s_{j}\psi_{j},\nonumber\\
\psi_{j}^{\left(  3\right)  }  &  =\ell_{j}^{-2}s_{j}\psi_{j}^{\prime}%
+\ell_{j}^{-3}\frac{V_{j}^{\prime}}{V_{z,j}}\psi_{j},\nonumber\\
\psi_{j}^{\left(  4\right)  }  &  =\ell_{j}^{-4}s_{j}^{2}\psi_{j}+2\ell
_{j}^{-3}\frac{V_{j}^{\prime}}{V_{z,j}}\psi_{j}^{\prime}+\ell_{j}^{-3}%
\frac{V_{j}^{\prime\prime}}{V_{z,j}}\psi_{j},
\end{align}
with $V_{z,j}=\left(  V_{j+1}-V_{j}\right)  /\Delta_{j}$, and obtain with the
expressions (\ref{tau}), (\ref{taylor}) and (\ref{airy3})%
\begin{equation}
\tau_{j+1}^{\psi}=\psi_{j+1}-\psi\left(  z_{j+1}\right)  =-\frac{1}{24}\left(
k_{j}^{2}\right)  ^{\prime\prime}\psi_{j}\Delta_{j}^{4}+\mathcal{O}\left(
\Delta_{j}^{5}\right)  ,
\end{equation}
and with (\ref{taus}), (\ref{taylors}) and (\ref{airy4})%
\begin{equation}
\tau_{j+1}^{\psi^{\prime}}=\psi_{j+1}^{\prime}-\psi^{\prime}\left(
z_{j+1}\right)  =-\frac{1}{12}\left(  k_{j}^{2}\right)  ^{\prime\prime}%
\psi_{j}\Delta_{j}^{3}+\mathcal{O}\left(  \Delta_{j}^{4}\right)  ,
\end{equation}
where $k_{j}=\sqrt{2m_{j}^{\ast}\left(  E-V_{j}\right)  }/\hbar$. Using
(\ref{airy2}), we can express the LDEs for the amplitudes $\mathcal{A}$ and
$\mathcal{B}$ in terms of $\tau_{j+1}^{\psi}$\ and $\tau_{j+1}^{\psi^{\prime}%
}$, and obtain $\tau_{j+1}^{\mathcal{A}}=\mathcal{O}\left(  \Delta_{j}%
^{3}\right)  $, $\tau_{j+1}^{\mathcal{B}}=\mathcal{O}\left(  \Delta_{j}%
^{3}\right)  $.

As described in Section \ref{sub:tra1}, a position dependent effective mass
can in the Airy function approach be treated by assuming a constant value
within each segment $j$, e.g., $m_{j}^{\ast}=m^{\ast}\left(  z_{j}\right)  $,
and applying the matching conditions (\ref{match}) at the section boundaries
\cite{1990IJQE...26.2025J}. The result for $\psi_{j+1}^{\prime}$ in
(\ref{airy4}) has thus to be multiplied by $m_{j+1}^{\ast}/m_{j}^{\ast}$
before inserting it into (\ref{taus}). While $\tau_{j+1}^{\psi^{\prime}}$ is
still $\mathcal{O}\left(  \Delta_{j}^{3}\right)  $, $\tau_{j+1}^{\psi}$ drops
to $\mathcal{O}\left(  \Delta_{j}^{2}\right)  $, now yielding $\tau
_{j+1}^{\mathcal{A}}=\mathcal{O}\left(  \Delta_{j}^{2}\right)  $, $\tau
_{j+1}^{\mathcal{B}}=\mathcal{O}\left(  \Delta_{j}^{2}\right)  $. The error
analysis also shows that $\tau_{j+1}^{\psi}$ and thus $\tau_{j+1}%
^{\mathcal{A}},\tau_{j+1}^{\mathcal{B}}$ can be improved to $\mathcal{O}%
\left(  \Delta_{j}^{3}\right)  $ by assigning an averaged mass $\left(
m_{j}^{\ast}+m_{j+1}^{\ast}\right)  /2$ rather than $m_{j}^{\ast}$\ to each
segment, and applying the matching conditions correspondingly.

\ifCLASSOPTIONcaptionsoff\newpage

\fi

\begin{IEEEbiography}
[{\includegraphics[width=1in,height=1.25in,clip,keepaspectratio]{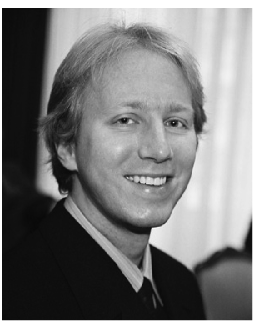}
}]{Christian Jirauschek} Christian Jirauschek was born in Karlsruhe, Germany,
in 1974. He received his Dipl-Ing. and doctoral degrees in electrical
engineering in 2000 and 2004, respectively, from the Universit\"{a}t Karlsruhe
(TH), Germany.

From 2002 to 2005, he was a Visiting Scientist at the
Massachusetts Institute of Technology (MIT), Cambridge, MA. Since 2005, he has
been with the TU M\"{u}nchen in Germany, first as a Postdoctoral Fellow and
since 2007 as the Head of an Independent Junior Research Group (Emmy-Noether
Program of the DFG). His research interests include modeling in the areas of
optics and device physics, especially the simulation of quantum devices and
mode-locked laser theory.

Dr. Jirauschek is a member of the IEEE, the German
Physical Society (DPG), and the Optical Society of America. Between 1997 and
2000, he held a scholarship from the German National Merit Foundation
(Studienstiftung des Deutschen Volkes).
\end{IEEEbiography}    
\end{document}